\input harvmac
%
 

\def\e#1{{\rm e}^{^{\textstyle#1}}}

\def\darr#1{\raise1.5ex\hbox{$\leftrightarrow$}\mkern-16.5mu #1}

\def\roughly#1{\raise.3ex\hbox{$#1$\kern-.75em\lower1ex\hbox{$\sim$}}}



\def\IB{\relax\hbox{$\inbar\kern-.3em{\rm B}$}}
\def\IC{\relax\hbox{$\inbar\kern-.3em{\rm C}$}}
\def\ID{\relax\hbox{$\inbar\kern-.3em{\rm D}$}}
\def\IE{\relax\hbox{$\inbar\kern-.3em{\rm E}$}}
\def\IF{\relax\hbox{$\inbar\kern-.3em{\rm F}$}}
\def\IG{\relax\hbox{$\inbar\kern-.3em{\rm G}$}}
\def\IGa{\relax\hbox{${\rm I}\kern-.18em\Gamma$}}
\def\IH{\relax{\rm I\kern-.18em H}}
\def\IK{\relax{\rm I\kern-.18em K}}
\def\IL{\relax{\rm I\kern-.18em L}}
\def\IP{\relax{\rm I\kern-.18em P}}
\def\IR{\relax{\rm I\kern-.18em R}}
\def\IZ{\relax\ifmmode\mathchoice
{\hbox{\cmss Z\kern-.4em Z}}{\hbox{\cmss Z\kern-.4em Z}}
{\lower.9pt\hbox{\cmsss Z\kern-.4em Z}}
{\lower1.2pt\hbox{\cmsss Z\kern-.4em Z}}\else{\cmss Z\kern-.4em Z}\fi}




\def\Tr{\rm Tr}



\def\unlockat{\catcode`\@=11}
\def\lockat{\catcode`\@=12}

\unlockat

\def\newsec#1{\global\advance\secno by1\message{(\the\secno. #1)}
\global\subsecno=0\global\subsubsecno=0\eqnres@t\noindent
{\bf\the\secno. #1}
\writetoca{{\secsym} {#1}}\par\nobreak\medskip\nobreak}
\global\newcount\subsecno \global\subsecno=0
\def\subsec#1{\global\advance\subsecno
by1\message{(\secsym\the\subsecno. #1)}
\ifnum\lastpenalty>9000\else\bigbreak\fi\global\subsubsecno=0
\noindent{\it\secsym\the\subsecno. #1}
\writetoca{\string\quad {\secsym\the\subsecno.} {#1}}
\par\nobreak\medskip\nobreak}
\global\newcount\subsubsecno \global\subsubsecno=0
\def\subsubsec#1{\global\advance\subsubsecno by1
\message{(\secsym\the\subsecno.\the\subsubsecno. #1)}
\ifnum\lastpenalty>9000\else\bigbreak\fi
\noindent\quad{\secsym\the\subsecno.\the\subsubsecno.}{#1}
\writetoca{\string\qquad{\secsym\the\subsecno.\the\subsubsecno.}{#1}}
\par\nobreak\medskip\nobreak}

\def\subsubseclab#1{\DefWarn#1\xdef
#1{\noexpand\hyperref{}{subsubsection}%
{\secsym\the\subsecno.\the\subsubsecno}%
{\secsym\the\subsecno.\the\subsubsecno}}%
\writedef{#1\leftbracket#1}\wrlabeL{#1=#1}}
\lockat

\def\IL{\relax{\rm I\kern-.18em L}}
\def\IH{\relax{\rm I\kern-.18em H}}
\def\IR{\relax{\rm I\kern-.18em R}}
\font\manual=manfnt \def\dbend{\lower3.5pt\hbox{\manual\char127}}

\def\c{\cdot}
\def\IZ{\relax\ifmmode\mathchoice
{\hbox{\cmss Z\kern-.4em Z}}{\hbox{\cmss Z\kern-.4em Z}}
{\lower.9pt\hbox{\cmsss Z\kern-.4em Z}}
{\lower1.2pt\hbox{\cmsss Z\kern-.4em Z}}\else{\cmss Z\kern-.4em
Z}\fi}


\def\IZ{\relax\ifmmode\mathchoice
{\hbox{\cmss Z\kern-.4em Z}}{\hbox{\cmss Z\kern-.4em Z}}
{\lower.9pt\hbox{\cmsss Z\kern-.4em Z}}
{\lower1.2pt\hbox{\cmsss Z\kern-.4em Z}}\else{\cmss Z\kern-.4em
Z}\fi}
\def\IB{\relax{\rm I\kern-.18em B}}
\def\IC{{\relax\hbox{$\inbar\kern-.3em{\rm C}$}}}
\def\ID{\relax{\rm I\kern-.18em D}}
\def\IE{\relax{\rm I\kern-.18em E}}
\def\IF{\relax{\rm I\kern-.18em F}}
\def\IG{\relax\hbox{$\inbar\kern-.3em{\rm G}$}}
\def\IGa{\relax\hbox{${\rm I}\kern-.18em\Gamma$}}
\def\IH{\relax{\rm I\kern-.18em H}}
\def\II{\relax{\rm I\kern-.18em I}}
\def\IK{\relax{\rm I\kern-.18em K}}
\def\IP{\relax{\rm I\kern-.18em P}}

\def\lies{{\underline{\bf s}}}
\def\inbar{\,\vrule height1.5ex width.4pt depth0pt}

\font\cmss=cmss10 \font\cmsss=cmss10 at 7pt
\def\IR{\relax{\rm I\kern-.18em R}}

\def\Tr{\rm Tr}


\def\boxit#1{\vbox{\hrule\hbox{\vrule\kern8pt
\vbox{\hbox{\kern8pt}\hbox{\vbox{#1}}\hbox{\kern8pt}}
\kern8pt\vrule}\hrule}}
\def\mathboxit#1{\vbox{\hrule\hbox{\vrule\kern8pt\vbox{\kern8pt
\hbox{$\displaystyle #1$}\kern8pt}\kern8pt\vrule}\hrule}}


\def\inbar{\,\vrule height1.5ex width.4pt depth0pt}

\font\cmss=cmss10 \font\cmsss=cmss10 at 7pt
\def\IR{\relax{\rm I\kern-.18em R}}

\def\Tr{\rm Tr}

%

  \lref\tonin{  G. Dall'Agata, K. Lechner and M. Tonin, {\it
Covariant actions for N=1, D=6 Supergravity theories with chiral bosons
}, hep-th/9710127,  Nucl. Phys. B512 (1998) 179-198.}

\lref\simons{  J. Cheeger and J. Simons, {\it Differential Characters and
Geometric Invariants},
 Stony Brook Preprint, (1973), unpublished.}

 \lref\seibergsix{  O. Aharony, M. Berkooz, N. Seiberg,  {\it Light-Cone
Description of (2,0) Superconformal Theories in Six Dimensions},
  hep-th/9712117\semi  O. J. Ganor, David R. Morrison, N. Seiberg,
 {\it
Branes, Calabi-Yau Spaces, and Toroidal Compactification of the N=1
Six-Dimensional $E_8$ Theory}, hep-th/9610251, Nucl. Phys. B487 (1997)
93-127\semi
N. Seiberg,
{\it Non-trivial Fixed Points of The Renormalization Group in Six
Dimensions},  hep-th/9609161, Phys. Lett. B390 (1997) 169-171.
}

 \lref\wittensix{E. Witten, {\it New  Gauge  Theories In Six Dimensions},
  hep-th/9710065.}

\lref\bks{  L.~Baulieu, H.~Kanno, I.~Singer,
{\it Special Quantum Field Theories in Eight and Other Dimensions},
hep-th/9704167, Talk given at
APCTP Winter School on Dualities in String Theory, (Sokcho, Korea),
February 24-28, 1997\semi
  L.~Baulieu, H.~Kanno, I.~Singer, {\it Cohomological Yang--Mills
Theory
in Eight Dimensions}, hep-th/9705127, to appear in Commun. Math.Phys.}

\lref\wittentopo { E. Witten,  {\it  Topological Quantum Field Theory},
Commun. Math.Phys. {117} (1988)353. }

\lref\wittentwist { E. Witten, {\it Supersymmetric Yang--Mills theory on a
four-manifold}, J. Math. Phys. {35} (1994) 5101.}

\lref\bs { L. Baulieu, I. M. Singer, {\it Topological Yang--Mills
Symmetry}, Nucl. Phys. Proc. Suppl.
15B (1988) 12\semi  L. Baulieu, {\it On the Symmetries of Topological Quantum
Field Theories},   hep-th/
9504015, Int. J. Mod. Phys. A10 (1995) 4483\semi R. Dijkgraaf, G. Moore,
 {\it Balanced Topological Field Theories},
hep-th/9608169,   Commun. Math. Phys. 185 (1997) 411.}

\lref\kyoto { L. Baulieu, {\it Field Antifield Duality, p-Form Gauge Fields
and Topological Quantum
Field Theories},     hep-th/9512026,   Nucl.Phys. B478 (1996) 431. }

\lref\strings {  L. Baulieu, M. B. Green, E. Rabinovici {\it A Unifying
Topological Action for Heterotic and  Type II Superstring  Theories},
hep-th/9606080, Phys.Lett. B386 (1996) 91\semi
{\it   Superstrings from   Theories with $N>1$ World Sheet Supersymmetry},
 hep-th/9611136, Nucl. Phys. B498 (1997). }

\lref\sourlas{  G. Parisi and N. Sourlas,
{\it Random Magnetic Fields, Supersymmetry and Negative Dimensions},  Phys.
Rev.Lett. 43 (1979) 744\semi  Nucl. Phys. B206 (1982) 321. }

\lref\SalamSezgin{A. Salam  and  E. Sezgin,
{\it Supergravities in diverse dimensions}, vol. 1, p.119\semi
P. Howe, G. Sierra and P. Townsend, Nucl Phys B221 (1983) 331}.

\lref\nekrasov{ A. Losev, G. Moore, N. Nekrasov, S. Shatashvili,
{\it
Four-Dimensional Avatars of Two-Dimensional RCFT},  hep-th/9509151,
Nucl. Phys. Proc. Suppl.  46 (1996) 130\semi L. Baulieu, A. Losev, N.
Nekrasov  {\it Chern-Simons and Twisted Supersymmetry in Higher
Dimensions},  hep-th/9707174, to appear in Nucl. Phys. B.  }

\lref\rabin {  L. Baulieu and  E. Rabinovici, in preparation.}

\Title{ \vbox{\baselineskip12pt\hbox{hep-th/9805200}
\hbox{CERN-TH:98-95}
\hbox{KCL-MTH:98-12}
\hbox{LPTHE:98-16 }}}
{\vbox{\centerline{   Six-Dimensional TQFTs   and
Twisted Supersymmetry}}}
\medskip
\centerline{Laurent Baulieu   }
\vskip 0.1cm
\centerline{CERN,
Geneva, Switzerland}
\vskip 0.1cm
\centerline{and}
\centerline{LPTHE\foot{UMR CNRS associ\'ee aux  Universit\'es Pierre et
Marie Curie (Paris VI) et Denis Diderot (Paris~VII).}, Paris, France.}

\vskip 0.5cm

 \medskip
\centerline{ Peter West   }
\vskip 0.1cm
\centerline{King's College, Strand,
London WC2 RLS, England.}\medskip
\vskip  1cm
\noindent
We describe a generalization of Yang--Mills topological field theory for
Abelian two-forms in six dimensions.   The connection of this theory
 by a twist to     Poincar\'e supersymmetric theories is given.  We also
briefly consider interactions and the case of self-dual three-forms in
eight dimensions.


  \Date{\ }

\newsec{Introduction}
In this paper  we introduce a generalization of topological field theory in six
dimensions for     two-form gauge fields. We analyse this theory using      the
techniques of BRST quantization for Topological Quantum Field Theories
(TQFT) thus   generalizing  to  six dimensions   the   Yang--Mills
TQFTs in four dimensions studied in
\wittentopo\wittentwist\bs\bks. In section 2
  we determine the   topological BRST operator $Q$  and
show that the result is related to    a    Poincar\'e supersymmetric
theory   in six dimensions with $(2,0)$
supersymmetry.  We     explain how $Q$ must  be combined with
the BRST
operator  of  the ordinary degenerate gauge symmetry of two-forms  to
obtain a completely gauge-fixed  action for the  TQFT.   We also briefly
comment on the Hamiltonian presentation of the TQFT, in the spirit of
\wittentopo.   In section 3, we show the possibility of introducing
interactions,
either by considering couplings of real two-forms to a Yang--Mills
field by a Chern-Simons term. We finish by noting that our results can be
generalized to  eight dimensions where the two-form is replaced by a
three-form and the corresponding twist leads to a supersymmetric
theory that contains a   gravitino, but no  graviton.

\def\e{\epsilon}
\def\demi{{1\over 2}}
\def\pa{\partial}
\def\a{\alpha}
\def\b{\beta}
\def\c{\gamma}
\def\m{\mu}
\def\n{\nu}
\def\r{\rho}

\def\L{L}
\def\X{X}
\def\V{V}
\def\M{m}
\def\N{n}
\def\H{H}
\def\Ic{{\it I}_{cl}}
\newsec{Free Case}

   \subsec{The fields and the $Q$-symmetry.}

Our aim is  to construct a TQFT in  six dimensions for
   two-form gauge fields   $B_2=\demi  B_{\mu\nu}dx^\mu   \wedge
dx ^\nu$. We chose the Minkowskian signature for the   six-dimensional space.
(Our formulae   can be adapted to    the Euclidian case, or other
signatures, with minor modification in the equations.)  Our  basic
hypothesis is that the TQFT
    explores quantum fluctuations around the solution of the self-duality
equations between two independant  real   two-forms  {
gauge fields} $B_2$ and  $^cB_2$
 \eqn\self{\eqalign{\pa_{[\mu} B_{\nu\rho ]}+ \e_{\mu\nu\rho\a\b\c
}\pa_{[\a} {^c B}_{\b\c ]}=0.}}
Having a pair of two-forms is necessary in order
to define a   ``topological''    invariant $\Ic$ that  does not vanish trivially
(the exterior product of a three-form by itself is zero).  We take for $\Ic$:
  \eqn\Icl{\eqalign{\Ic = \int dB_2\wedge d{^cB}_2.}}
This invariant can be  non trivial. Indeed,   $B_2$ and ${^c B}_2$ can be
defined as ``connections''    rather than as forms, and such   that
 $dB_2$ and $d{^c B}_2$ are still forms and thus are integrable.  (See \simons). Then, the
integral \Icl\ is well-defined, but can be non zero. Actually, up to a
normalization,   \Icl\ must be  ${\cal Z}$-valued. This may appear
as a motivation for the construction for a TQFT for two-forms in six
dimensions.

We notice  that the gauge condition \self\ implies self-duality and
antiself-duality relations  for the curvatures of $B_2+{^cB}_2$ and
$B_2-{^cB}_2$ respectively. As such, it imposes
for ten independent conditions  separately for each gauge field
combination.
 We will use the left hand side of \self\ as gauge functions  for the BRST-invariant gauge-fixing of $\Ic$,   since
the number of  gauge-independent components of the  two-forms $B_2$ and  $^cB_2$
in six dimensions is precisely
$2C^2_5 =20$.

 The  theory  must exhibit a topological BRST symmetry with a generator $Q$.
$Q$   is
 a graded differential operator acting  on  the gauge-invariant sector of
the TQFT,   whose     fields are defined in the following expansion (the
upper index is the ghost number) \eqn\fieldtop{\eqalign{& F_{\mu \nu\rho}
dx^\mu   \wedge dx ^\nu   \wedge dx ^ \rho
+\Psi_{\mu\nu}^1dx^\m   \wedge dx ^\n  +   \Psi_{\mu\nu}^{-1}dx^\m   \wedge
dx ^\n  \cr
&+\Phi_\m^{2}dx^\m+ \Phi_\m^{-2}dx^\m  +\L_\m ^{ }dx^\m\
+\Phi ^{3} +\X ^{1}+ \X ^{-1}+ \Phi ^{-3}.
\cr}  }
 The other two-form $^cB_2$   possess  an identical
expansion \fieldtop\ and the foregoing  equations must be duplicated
in an obvious way.

 A geometrical signification  presumably exists (as suggested by the
formula (2.29-2.30) of section 2.4) for
the ghosts
$\Psi^1_{\m\n}, \Phi_\m^2, \Phi^3$. For these fields, one has:
\eqn\symQf{\eqalign{& Q B_{\mu \nu} = \Psi_{\mu\nu}^1  \quad  Q
\Psi_{\mu\nu} ^1=\pa_{[\m}\Phi^2_{\n]} \cr
& Q \Phi_{\mu  }^2 = \pa_{\mu }  \Phi^3\quad  Q \Phi^3=0.
 \cr}  }
$Q$   is nilpotent up to a (degenerate) gauge transformation of $B_2$, with
``parameter''    $\Phi^2_\m$. It can be used to define the equivariant cohomology of two-forms, modulo their gauge transformations which are discussed in section 2.4. 

 The  rest of the  fields in  \fieldtop\ (not including the classical
curvature $F_{\mu \nu\rho}$ of the two-form $B_{\mu\nu}$) are antighosts,
which are   related to   antifields in the Batalin--Vilkoviski formalism
(see for instance  \kyoto) and transform into   auxiliary (Lagrange
multiplier) fields:
\eqn\symQaf{\eqalign{& Q  \Psi_{\mu \nu}^{-1} = \H_{\mu\nu} \quad  Q
\H_{\mu\nu} =0
\quad \quad   Q  \Phi_{\mu  }^{-2} = \eta^{-1}_\m  \quad  Q \eta^{-1}_\m =0
 \cr
& Q  \L_{\mu  }  ^{ }= \eta^1_\m \quad    \quad \quad  Q  \eta_\m^1 =0
\quad \quad    Q  \Phi^{-3}  = \eta^{-2}\quad  Q  \eta^{-2} =0
 \cr
& Q  \X^{-1}  = \eta ^{ }\quad  \quad  \quad Q  \eta ^{ } =0
\quad \quad \quad  Q  \X^{ 1}  = \eta^{2}\quad  Q  \eta^{2} =0.
 \cr
}  }

Our purpose is to  define a theory whose topological gauge conditions are  the
20=10+10     self-duality conditions for
$B_2$
 and   $^c B_2$. To do so, we need to introduce 10 self-dual and
10
antiself-dual three-form Lagrange multipliers. These must be defined
from  the 15+15 fields
$H_{\m\n}$ and $^c H_{\m\n}$
Following the spirit of
\bs, this leads us to decompose  the two-form antighosts
$\Psi_{\mu\nu} ^{-1}$ with fifteen components into a self-dual three-form
$\chi_{\m\n\r} ^{-1} $ with ten components
 and a field $\chi_ L^{-1}$  with five components  $\chi_{L(a)}^{-1}$, where
$a$ runs from $1$ to $5$. (This decomposition is not
Lorentz covariant at each step, but
the result is   the Lorentz covariant BRST
multiplet (2.9).)   The same is done  for the Lagrange multipliers
that are the BRST transforms of the antighosts, which gives
$\H_{\m\n\r}  $   and   $\H_{L} $, with
$\H_{\m\n\r} =
\big\{Q,  \chi_{\m\n\r} ^{-1} \big\} $   and   $\H_{L} =
\big\{Q,  \chi_{L} ^{-1} \big\} $.  For the  second set of fields with
upper index $^c$,
we  take
$^c\chi_{\m\n\r} ^{-1} $ and $^c \H_{\m\n\r}   $ as   antiself-dual
three-forms.

The   antighosts and Lagrange multipliers
$\X^{-1},\chi_{L(a)}^{-1 }, \eta ^{1}_\mu, \eta,\H_{L(a)}^{{} }$ and $ \L_\mu$
must be eliminated from the TQFT. This can be done by a brute-force
gauge-fixing, obtained from  the following
$Q$-exact Lagrangian:
\eqn\gf{\eqalign{  &\big\{Q, \   \X^{-1}\L^{}_6  +\
\sum_{a=1}^5
\chi_{L(a)} ^{-1  }\L^{{} } _{a} \big\}=
   \eta ^{ }\L^{}_6 - \X^{-1}\eta^{1 }_6
+\sum_{a=1}^5
(\H_{L(a)}^{{} } \L^{}_{a}
-\chi_{L(a)}^{-1 }\eta^{1 } _{a} ).
 \cr
}  }
  The      equations of motion are delta functions that provide the desired
gauge-fixing to zero of the  fields
$\X^{-1},\chi_{L(a)}^{-1 }, \eta ^{1}_\mu, \eta,\H_{L(a)}^{{} }$ and $
\L_\mu$.
Their     elimination     can be rephrased as follows:   the $-5$
degrees of freedom of the fermions    $\chi_{L(a)}^{-1 }$ are compensated by the
$+6$ degrees of freedom of  the  bosons $\L^{}_\m$ plus the $-1$  degree of
freedom
of the fermion  $  \X^ {-1}$; said differently,
  a self-dual three-form
$\chi^{-1}_{\mu \nu\rho}$ in six dimensions is   the
combination of    the  fields $( \Psi_{\m\n}^{-1}, L_\m^{},   X^{-1})$
with  algebraic  compensations between commuting and anticommuting  fields.

We  thus reduce the expansion  \fieldtop\  and the definition \symQaf\ of
$Q$   as follows \eqn\fieldtopr{\eqalign{  F_{\mu \nu\rho} dx^\mu   \wedge
dx ^\nu   \wedge dx ^ \rho
+\Psi_{\mu\nu}^1dx^\m   \wedge dx ^\n  +\chi^{-1}_{\mu \nu\rho} dx^\mu
\wedge dx ^\nu   \wedge dx ^ \rho  \cr  +\Phi_\m^{2}dx^\m+
\Phi_\m^{-2}dx^\m   +\Phi ^{3} +\X ^{1} + \Phi ^{-3}
\cr}  }
\eqn\symQf{\eqalign{  Q B_{\mu \nu} = \Psi^1_{\mu\nu} \quad  Q
\Psi^1_{\mu\nu} =\pa_{[\m}\Phi^2_{\n]}
\quad
Q \Phi_{\mu  }^2 = \pa_{\mu } \Phi^3\quad  Q \Phi^3=0
 \cr
  Q  \chi_{\mu \nu\r}^{-1} = \H_{\mu\nu\r} ^{ } \quad  Q \H_{\mu\nu\r} ^{ } =0
\quad
\quad
Q  \Phi_{\mu  }^{-2} = \eta^{-1}_\m  \quad  Q \eta^{-1}_\m =0
 \cr
   Q  \X^{ 1}  = \eta^{2}\quad  Q  \eta^{2} =0
\quad
Q  \Phi^{-3}  = \eta^{-2}\quad  Q  \eta^{-2} =0.
 \cr
}  }

The bosonic fields  $H_{\m\n\r}$ and $\eta^{\pm 2}$ will eventually be
eliminated as
Lagrange multipliers for gauge conditions.  Therefore,
 the ``basic"
TQFT multiplet  for a   two-form with self-duality gauge conditions is
\eqn\mul {\eqalign{(B_{\m\n}, \Psi_{\m \n }^{1}, \chi_{\mu \nu\r}^{-1},
\eta_{\mu  }^{-1} ,\Phi^{ 3},   \Phi^{ -3}, \X^{1}, \Phi^{ 2}_\m,  \Phi^{
-2}_\m).
\cr }  }
This multiplet  could have been our starting point, in particular
in the  Hamiltonian construction   sketched in section 3.  Since we
have in addition the fields arising from the gauge field
${}^cB_{\mu\nu}$ we have in effect twice  the number of the
above fields in the actual topological theory. Let us again stress that in
\mul,
$B_{\m\n}$ is unconstrained. The TQFT multiplet of   $^c B_{\m\n}$  is
completely analogous to \mul, except that
 $^c \chi_{\mu
\nu\r}^{-1}$ is antiself-dual  rather than  self-dual.

\subsec{Twist.}
The multiplet  \mul\  contains  $34=15+10+6+3$  fermionic
degrees of freedom   and $15+12 $
 bosonic   degrees of freedom. To make the link to Poincar\'e
supersymmetry,   we
assume that two bosonic and two  fermionic degrees  of freedom, e.g
$\Phi^{ \pm 3}$ and $\Phi^{ \pm 2}_6$,
  compensate  each other  by supersymmetry, as two topological packages in the
language of \strings. What remains is a system of
 $32$ fermions and $15+ 10$ bosons, which we write as
 \eqn\mulr {\eqalign{(B_{\m\n}, \Psi_{\m \n }^{1},
\chi_{\mu \nu\r}^{-1},  \eta_{\mu  }^{-1}, \X^{1}, \Phi^a),\cr}  }
with  $1\leq a\leq 10$.

We now wish to find a supersymmetric multiplet that,
after a twist, can be identified with the above fields. Since
the fields in the above equation  do not include a graviton this
supersymmetric multiplet must possess supersymmetry transformations
with  less than 32 supercharges. As we are in six dimensions the
largest possible number of supercharges that is less than 32 is 16.
The supermultiplet with this number of supercharges and  a second-rank
antisymmetric tensor gauge field is
$(2,0)$ tensor multiplet
\SalamSezgin.

Before giving the details of the twist in six dimensions
it will be instructive to  recall some of the essential features
of the twist to relate the topological Yang--Mills theory in four
dimensions to the supersymmetry multiplet of $N=2$
supersymmetric Yang--Mills theory \wittentwist. The
$N=2, D=4$ supersymmetric  multiplet is given by
$(A_m, \lambda_{\a}^{i},   \bar \lambda_{\dot\a i},B,\bar B)$ and it
is related by the twist to  the TQFT  multiplet $(A_m, \Psi_m^{(1)},
\chi_{m n  } ^{(-1)},
\eta ^{(-1)},\Phi ^{(2)} ,  \Phi ^{(-2)})$ in the notation of
ref.
\wittentwist. The
essential step in the twist proceedure is the identification of the
$SU(2)_{susy}$ and $SU(2)_{R}$ groups in the
$SO(4)\times SU_{susy} (2)\sim  SU_L(2)\times SU(2)_R\times
SU(2)_{susy}$ invariance of the SSYM theory \wittentopo\wittentwist.

After setting the supersymmetry parameter to
$\eta ^{\alpha i}=0, \bar \eta ^{\dot \alpha\dot  \beta }=
-\epsilon ^{\dot \alpha\dot  \beta }\rho$  we find
\wittentwist\ that the $N=2$
Yang--Mills transformations for $A_\m$, $\lambda_{\a \dot \beta }=
(\sigma^m)_{\a \dot \beta }\Psi_m^{(1)}$ and $\Phi^{(2)}\equiv B$
give  the correct transformations for the original gauge fields and the
ghosts.  However, this discussion can be  slightly extended  to
 recover also  the BRST transformations of the anti-ghosts and Lagrange
multipliers. Indeed, making the identifications
$ \bar \lambda_{\dot\a \dot \beta }\equiv (\sigma^{mn} )_{\dot\a \dot
\beta }\chi_{mn}^{(-1)}+\epsilon _{\dot\a \dot
\beta } \eta ^{(-1)}$ and $\bar B\equiv \Phi^{(-2)}$, we
find that the  supersymmetry transformations of eq. (3.4) of
ref.
\wittentwist\ imply, up to constants, that
$\delta \chi_{mn}^{(-1)}= \rho (F_{mn}-{1\over
2}\epsilon_{mnpq}F^{pq}),
\delta \Phi^{(-2)}=\rho\eta^{(-1)}$ and $\delta \eta ^{(-1)}=0$. These
variations
are the BRST transformations  of the antighosts after
the self-dual Lagrange multiplier has been
eliminated  to enforce the gauge
condition and the auxiliary field in the $N=2$ Yang--Mills theory
has been set to  vanish.

Let us now consider  how   the fields and the transformation laws of   the
BRST symmetry
\symQf\ are  related to those of the (2,0) tensor multiplet
of Poincar\'e supersymmetry in  six dimensions.
\def\spinor{\kappa}
\def\scalar{\Phi}
The six-dimensional spinors we  are interested in transform
 under   $Spin(1,5)\times$
$Spin(5)$. The first group in this direct product is
by definition the  group
under which a six-dimensional spinor transforms. The second group
is an internal group that arises naturally for six-dimensional
spinors that have their origin  in eleven dimensions,  where they
transform under
$Spin(1,10)$. In fact,
$Spin(5)$ is isomorphic to
$USP (4)$.  Spinors in six  dimensions have $2^3=8$ components. The
matrix $B$, which implements the relation $B\gamma^\mu B^{-1}
=\gamma^{\mu^*}$ satisfies, $B^\dagger B=1$ and $\tilde B=-B$. As in
any even-dimensional space we can have Weyl spinors,
 but the
properties of $B$ imply that Majorana spinors (i.e.
$\psi^* = B\psi$)  do not exist  in six dimensions. We
work here not with the full $8\times 8$ six-dimensional
$\gamma $-matrices, but with their chirally projected components
which are $4\times 4$ matrices.
\par
However, in six dimensions we can  have symplectic
Majorana--Weyl  spinors, which have the indices $\spinor_{\alpha i}$ where
the indices $\alpha=1,\dots,4$  transform under
 Weyl-projected $Spin(1,5)$ transformations   and
the indices $i=1,\dots,4$
transform under $ Spin (5)$ transformations. The Majorana--Weyl
condition reads
 \eqn\weyl{\eqalign{
\spinor_{\alpha i} ^*=B_{\dot\alpha}^{\ \beta}\Omega^{ij}
\spinor_{\beta j},
}}
 where
$\Omega^{ij}$ is the antisymmetric real metric of $USP (4)$, which
raises and lowers indices by
$S^i\equiv\Omega^{ij}S_j$ and $S_j=S^k\Omega_{kj}$.
The matrix $B$ is block diagonal.  If we denote its
upper- and lower-blocks by
$B_{\dot\alpha}^{\ \beta}$ and
$B_{\dot\gamma}^{\ \beta}$ respectively, we can convert
dotted indices  to undotted ones (i.e.
$\chi_{\dot\beta}=B_{\dot\beta}^\gamma\chi_\gamma$ and
$\chi^\beta=\chi^{\dot\gamma}B_{\dot\gamma}^{\ \beta}$). We note,
however, that we cannot raise and lower the spinor
indices using the $B$ matrix.

The simplest supermultiplet that contains such a Majorana--Weyl spinor is
the so-called $(2,0)$ tensor
multiplet which possess the real fields $\scalar_{ij},\
B_{\mu\nu}^{}$ and spinors
$\spinor_{\alpha i}$. The ``on-shell''    counting of degrees of freedom
goes as follows: the  $\spinor_{\alpha i}$
obey   eq. \weyl\ and so contribute  $8$ on-shell
 degrees of freedom; the five
scalars live in the real field
$\scalar_{ij}$, which obeys the constraints
\eqn\trace{\eqalign{
\scalar_{ij}=-\scalar_{ji},\quad\Omega^{ij}\scalar_{ij}=0.
}}
The    gauge field $B_{\mu\nu}^{}$ possesses a self-dual
field strength $F_{\rho\mu\nu}^{(+)}=3\partial_{[\rho}
B_{\mu\nu]}^{}$ (i.e. $F_{\rho\mu\nu}^{(+)}={1\over3!}
\varepsilon_{\rho\mu\nu\spinor\tau\kappa}F^{\spinor\tau\kappa (+)}$),
which provides only $3$ on-shell degrees of freedom.
 This is the multiplet
that arises in the 5-brane of M-theory. This (2,0) supermultiplet
transforms under  supersymmetry with a parameter that is also
Majorana--Weyl and so has the desired 16 components.

The supersymmetry variations of the component fields
of the (2,0) tensor multiplet are given, up to
constants, by

\def\eb{{\bar \epsilon}}
\eqn\susytr {\eqalign{ &
\delta \kappa _{\a i} =  F_{\m\n\r}^{(+)}(\gamma^{\m\n\r})_{\a
\b}\e^\b_i+ (\gamma^\m\partial_\m)_{\a \dot\b}\Phi _{ij} \eb^{\dot\b
j}\cr &
\delta \bar \kappa _{\dot \alpha }^{ i} =
\bar F_{\m\n\r}^{(+)}(\gamma^{\m\n\r})_{\dot\a \dot\b}\eb^{\dot \b i}+
(\gamma^\m\partial_\m)_{\dot \a  \b}\bar \Phi ^{ij} \e^{\b j}  \cr &
\delta B _{\m\n }^{}  =  \e^\a_i(\gamma_{\m\n})_{ \a }^{\dot\b}
\bar  \kappa^i_{\dot \b} \cr
&\delta \Phi_{ij} =
\e^\b_{[i} \kappa_{ \b j]} -{1\over 4} \Omega _{ij}
 \epsilon ^\b_k\kappa ^k_\beta
\cr
&\delta \bar \Phi^{ij}
= \bar  \kappa_{ \dot\a }^{[i}  \eb^{\a j]} +{1\over 4} \Omega ^{ij}
 \bar \kappa ^k_{\dot \a}\eb _k^{\dot \a}.
\cr
}}

Examining
the variation of the spinors,  we find
that the self-duality condition is associated with the Weyl nature of
the spinors as the Weyl projected $\gamma$-matrix $(\gamma
^{abc})_{\alpha\beta}$ is automatically antiself-dual. On the other
hand, the reality condition of the scalars and the gauge field
in the (2,0) supermultiplet follows from the symplectic Majorana
condition of eq.
\weyl\ when  imposed on the supersymmetry
variation of the spinors. Of course $\kappa $ and $\bar \kappa$ are
related by their Majorana--Weyl condition, while  $\Phi_{ij}$ and
$\bar \Phi^{ij}$ are related by lowering their indices, but
it will be convenient to  write the transformation laws in this way.

\par

We now wish to twist the above supersymmetry transformations
and recover
the topological BRST transformations. Clearly the six-dimensional
spin group $Spin(1,5)$ contains
$Spin(5)$. The twist consists in identifying these $Spin(5) $
transformations with those of the internal $Spin(5) $ symmetry group.
 The
group $USP (4)$ consists of unitary matrices that also preserve
$\Omega^{ij}$, i.e. matrices  $A$ that satisfy $A^TA=1$ and
$A^T\Omega A=\Omega$. The latter condition can also be written as
$\Omega A=A^*\Omega$.
On the other hand the $Spin(1,5)$ acts separately on each chiral
sector and the chirally projected $Spin(1,5)$ transformations are
isomorphic to $SU^*(4)$, which is the group of four by four complex matrices
of determinant $1$, which obey $BA=A^*B$. Hence to identify the two
$Spin(5)=USP (4)$ groups also requires us to identify $\Omega=B$.

Having identified the two $Spin(5)$ groups,  we may place
$Spin(5)
$ invariant conditions on the supersymmetry parameters
$\varepsilon^\alpha_{\ i}$ and $\bar\varepsilon^{\dot\alpha i}$.
 In
particular, we choose
\eqn\cond{\eqalign{
\bar\varepsilon^{\dot\alpha i}
=0,\quad\varepsilon^\alpha_{\ i}=\delta^\alpha_i\rho.
}}
We may write the latter constraint as
$\varepsilon^{\alpha i}=\rho\Omega^{i\alpha}$ by raising indices. We
can now identify $i,j\dots$ indices with $\alpha,\beta,\dots$ indices,
since they transform in the same way under the $Spin(5) $
identification. The Grassmann odd parameter
$\rho$ will be the  ``infinitesimal    parameter''    of the BRST transformation.
We note that, the twist
breaks the Majorana--Weyl condition on the supersymmetry parameters
$\varepsilon^\alpha_{\ i}$ and $\bar\varepsilon^{\dot\alpha i}$, as indeed it also breaks the    Majorana  condition in four dimensions in \wittentwist.
Moreover, just as we dropped the Majorana--Weyl constraint on the
supersymmetry parameter, we also drop it on the spinors of the
multiplet, and we regard the scalars 
$\scalar^{ij}$ and $\bar \scalar^{ij}$ as independent fields.
\par
Upon substituting the choice of supersymmetry parameters \cond,
the transformations \susytr\   become:
\eqn\susytra {\eqalign{
&\quad\quad\quad \quad\quad\quad\quad
\delta B _{\m\n }^{}  = \r  (\gamma_{\m\n})_{ \a }^{\ \dot\b}
\bar \kappa^\a_{\dot \b}
 \cr
 &
\delta \kappa _{\a \b} =  \r F_{\m\n\r}^{(+)}(\gamma^{\m\n\r})_{\a \b}
\quad\quad
\delta \bar \kappa _{\dot \a }^{ \gamma} =  \r
(\gamma^\m\partial_\m)_{\dot \a  \b}\bar \Phi ^{\gamma\beta}
 \cr
 &\delta \Phi_{ij}
=\r
  \kappa_{ ij}-{1\over 4} \Omega _{ij}\rho \kappa_k^k
 \quad\quad\quad
\quad\quad\quad\quad
\delta \bar \Phi_{ij}
=0.\cr
}}

We can expand  the  spinors $\spinor_{\alpha
i}\equiv\spinor_{\alpha\beta}$ and
$\bar\spinor_{\dot\alpha}^{\ i}
\equiv\spinor_{\dot \alpha} ^{\ \beta}$
 in terms of the complete set
of chiral
$\gamma$-matrices as follows\eqn\cor {\eqalign{ &
\kappa_{\a\b}=  (\gamma^{\m\n\r})_{\a \b}\ \chi^{-1 (+)}_{\m\n\r}
+(\gamma^{\m })_{\a \b}\ \eta^{-1}_{\m }
\cr
&
\kappa_{\dot \a}^{\b}=  (\gamma^{\m\n})_{\dot \a}^{\b}\
\Psi^{1 }_{\m\n }
+  \delta _{\dot \a}^{\b}\
\X^{-1},
\cr
}  }
which  makes precise
 the correspondence    between the fermions of the TQFT
and those of the supersymmetric theory. The matrices
$(\gamma^\mu)_{\alpha\beta}$ are antisymmetric in
$\alpha\beta$ while
$(\gamma^{\mu\nu\rho})_{\alpha\beta}$ are   symmetric in
$\alpha\beta$ and also antiself-dual. As a result,
 $\chi_{\m\n\r}^{-1(+)}$ is automatically a self-dual
three-form.
 On the other hand the   scalars in
$\scalar^{ij}\sim \scalar_{\alpha\beta}$ can be written as
\eqn\coranti {\eqalign{ &
\bar \Phi_{\a\b}=  (\gamma^{\m })_{\a \b}\Phi^{2}_\m
\quad\quad
  \Phi^{\a\b}=  (\gamma^{\m })^{\a \b}\Phi^{-2}_\m.\cr
}  }

We must also impose the condition
$\Phi^{\alpha\beta}\Omega_{\alpha\beta}=0$
and hence there are only ten independent components in  $\Phi^{\pm 2}_\mu$. This
condition  
implies that the  TQFT multiplet contains two more bosons than the
supersymmetric one, as indicated by
\coranti. This is compensated by the presence of the two extra
fermions
$\Phi^{\pm 3}$. (See next section.) \par
 Let us now combine the  symmetry \susytra\ and the
redefinitions \cor\ and \coranti, so as  to find the transformations
\eqn\susytransf {\eqalign{
\delta B _{\m\n }  =\r  \Psi^{1}_{\m\n }
 \quad
\quad \delta \Phi^{2}_{ \m  } &=0
\quad
\quad\delta     \eta_\m  ^{-1} =0
 \cr
\delta \Psi^{1}_{\m\n } =  \r  \partial_{[\mu} \Phi^{2}_{ \n] }
 \quad
\quad
 \delta \X  ^{1}   &=  \r \partial _\m \Phi^{2\m} \quad
\quad
  \delta \Phi  ^{-2}_{\m } =  \r \eta_\m
^{-1}\cr
  \delta \chi  ^{-1(+)}_{\m\n\r}  &=   \r   F_{\m\n\r} ^{(+)}\quad\quad.
}}
In these equations  the indices on the fields  $\Phi^{\pm 2}_{\m}$
only take the values $\mu=1,2,\ldots 5$ and similarly for
$\eta_\mu$, in order to take into account \coranti.

The comparison to the BRST equations \symQf\ also  requires $\Phi^{\pm 3} $   to be   set to $0$
  and  the Lagrange multiplier  field
$H_{\m\n\r}$    to  be replaced     by the  self-duality   gauge
function for
  $B_2$ and $\eta^{\pm 2}$     by $\pa_\m \Phi^{\m \pm
2}$. The  latter conditions appear as equations of motion
in the
$Q$-invariant action that we will consider in the next section (see eqs. (2.21) and (2.24)).

In addition to the above,  we must introduce an analogous
supersymmetry multiplet and carry out a corresponding twist to
reproduce the BRST transformations of the  another two-form
${^c B}_2$,      with similar steps as we did for $B_2$ and its ghosts.

 \subsec{The     $Q$-invariant and gauge-invariant action.}
 The property $Q^2=0$    modulo  gauge  transformations  of    the 
two-form $
B_2$
 leads us to consider
  the  following $Q$-invariant and  gauge-invariant Lagrangian
(the gauge parameter $\a$ is a real number, which gives the self-duality
condition \self\
in the limit
$\a\to 0$):
 \def\top{Q }
\eqn\lsymQf{\eqalign{ I_{ \top}  =\int  \ \big\{ Q, \  \chi^{\m\n\r} (
\pa_\mu B_{\nu\rho}+\e_{\mu\nu\rho\a\b\c }\pa_\a {^c B}_{\b\c}
 +
 \a H_{\m\n\r})
\cr
   \Phi^{-2}_\m \pa_\n\Psi^1_{[\m\n]}+
  \Phi^{-3}\pa_\m\Phi^2_\m
 +\X^1(  \eta^{-2}+\pa_\m  \Phi_\m^{-2}) \    -cc \big\}.
}  }
In this expression and the foregoing, the notation $cc$ means
  terms identical to the
  explicit ones, which are obtained by replacing all fields with their
counterparts with the
index
$^c$. (One must remember that $H_{\m\n\r}$ and $^cH_{\m\n\r}$, as well as
$\chi_{\m\n\r}$ and $^c\chi_{\m\n\r}$,
 are
respectively  self-dual and antiself-dual).
The action
\lsymQf\ indicates our choices for the gauge functions of ghosts and antighosts
\eqn\clever{\eqalign{&
\pa_\n\Psi^1_{[\m\n]}\cr &   \pa_\m\Phi^2_\m\cr &\pa_\m \Phi_\m^{-2}.
} }
Using the definition of $Q$ and eliminating the   fields $H_{\m\n\r}$ and
$\eta^{\pm2 }$ by their algebraic equations of motion
\eqn\motion{\eqalign{ &H_{\m\n\r}=\a^{-1}(\pa_\mu
B_{\nu\rho}+\e_{\mu\nu\rho\a\b\c}
\pa_\a {^cB}_{\b\c})^{(+)}
\quad\quad \eta^{\pm 2}=\pa_\m \Phi_\m^{\pm2},
\cr
}  }
and since:
 \eqn\aie{\eqalign{\Ic+\int (|\pa_\mu B_{\nu\rho}+\e_{\mu\nu\rho\a\b\c }\pa_\a
{^c B}_{\b\c}|^2-cc)
= \int( |\pa_\mu B_{\nu\rho}|^2
-|\pa_\mu {^cB}_{\nu\rho}|^2 ),
}}
we get
\eqn\symQfc{\eqalign{ \Ic+I_{\top} \sim \int   &
\a^{-1}
|\pa_\mu B_{\nu\rho}|^2
+ \chi^{\m\n\r}
\pa_{[\mu} \Psi^1_{\nu\rho]}
+\eta_\m ^{-1}\pa_\n\Psi_{[\m\n]}^{1}
+\X^1\pa_\m\eta_\m^{-1}
+  \Phi^{-3}\pa_\m\pa^\m \Phi^{ 3}
\cr
&
+  \Phi^{-2}_\m \pa_{ \n}\pa_{[\m }\Phi^{ 2}_{\n ]}
+\pa_\n   \Phi_\n^{-2}\pa_\m \Phi^{ 2}_\m
-cc.
\cr
}  }
The BRST invariance of the action, after   elimination of auxiliary fields, is
\eqn\symQafter{\eqalign{  Q B_{\mu \nu} = \Psi^1_{\mu\nu} \quad  Q
\Psi^1_{\mu\nu} =\pa_{[\m}\Phi^2_{\n]}
\quad
Q \Phi_{\mu  }^2 = \pa_{\mu } \Phi^3\quad  Q \Phi^3=0
 \cr
  Q  \chi_{\mu \nu\r}^{-1} = \a^{-1}(F_3+^{*c}F_3)^{(+)}_{\mu\nu\r}
\quad
\quad
Q  \Phi_{\mu  }^{-2} = \eta^{-1}_\m  \quad  Q \eta^{-1}_\m =0
 \cr
   Q  \X^{ 1}  = \pa_\m \Phi_\m^{2}\quad
\quad
Q  \Phi^{-3}  = \pa_\m \Phi_\m^{-2}.
 \cr
}  }
One has of course  all mirror equations obtained  by putting   the index
$^c$ on all fields, and  $ Q  ^c{\chi}_{\mu \nu\r}^{-1} =\a^{-1}
(F_3+^{*c}F_3)^{(-)}_{\mu\nu\r}$.

The two last terms of  the action \symQfc\ can be combined into
$    \Phi_\n^{-2}\pa^\m \pa_\m \Phi^{ 2}_\n$. Thus, one has finally:
 \eqn\symQfcc{\eqalign{
 \Ic+I_{\top} \sim
I_{ top} = \int   &
\a^{-1}
|\pa_\mu B_{\nu\rho}|^2
- \chi^{\m\n\r}
\pa_{[\mu} \Psi^1_{\nu\rho]}
+\eta_\m ^{-1}\pa_\n\Psi_{[\m\n]}^{1}
-\X^1\pa_\m\eta_\m^{-1}
\cr
&
-  \Phi^{-3}\pa_\m\pa^\m \Phi^{ 3}
+  \Phi_\n^{-2}\pa^\m \pa_\m \Phi^{ 2}_\n -cc.
\cr
}  }

This  $SO(5,1)$-invariant  $Q$-invariant action, which we have determined
by using  the self-duality
gauge function in\self\ for the two-form   and Feynman-Landau-type  gauge
functions for     the  gauge
invariances of  the topological ghosts,  is suitable for defining the TQFT.

To uncover the    relation to a   supersymmetric action,     one must
cancel the fermionic ghosts $\Phi^{\pm 3}$   against, for instance, the components
$\Phi^{\pm 2}_6$ of the vector ghosts
 $\Phi^{\pm 2}_\m$, as already    indicated in  section 2.2 at the level of
transformation laws. This gives  the action
\eqn\symQfcct{\eqalign{
  \int   &
\a^{-1}
|\pa_\mu B_{\nu\rho}|^2
- \chi^{\m\n\r}
\pa_{[\mu} \Psi^1_{\nu\rho]}
+\eta_\m ^{-1}\pa_\n\Psi_{[\m\n]}^{1}
-\X^1\pa_\m\eta_\m^{-1}
  +\sum _{a=1}^{5}  \Phi_a^{-2}\pa^\m \pa_\m \Phi^{ 2}_a -cc.
\cr
}  }
The
compensation   between  $\Phi^{\pm 3}$   and    $\Phi^{\pm 2}_6$,  which we
can call ``topological packages''     as in \strings, can in principle be done
in a Parisi--Sourlas-type way \sourlas.  The $SO(1,5)$ symmetry now seems
broken
down to $SO(5)\sim   USP(4)$; as shown in section 2.2, the restoration of
$SO(1,5)$ invariance in the context of $(2,0)$ $D=6$ supersymmetry can be
achieved by     reinterpreting      the  ten fields
$\Phi^{\pm 2}_a $,   $ 1\leq a\leq 5$,  as the ten
$SO(1,5)$ scalar  components of a complex
    traceless antisymmetric $USP(4)$  tensors $\Phi_{ij}$: then,
  by  doing all other changes of variables    \cor\ and  \coranti, the
action \symQfcct\   reads as  a    sum of  two free six-dimensional actions
depending on     a self-dual  and an antiself-dual (complex) two-form.

 Our conclusion is      that  the free TQFT action
 \symQfc\ can be twisted into    a free $SO(5,1)$ invariant action depending on  a pair of $D=6$ Poincar\'e   two-form multiplets
plus    a decoupled  action depending on  two compensating
``topological packages".  Overall, this  mechanism is reminescent  of that found
in
\strings, where  all superstring
actions  are  connected, by various twists, to a
purely topological unifying superstring action with a large enough value of
supersymmetry on the worldsheet.


\subsec{Completion of the gauge-fixing. }
We now turn to the gauge-fixing of the remaining   ordinary gauge
invariances of the action \symQfcc. The  result must be a completely
gauge-fixed action, invariant under a completely nilpotent topological BRST
symmetry.
The ordinary ghost field spectrum   of a two-form is
\eqn\field {\eqalign{& B_{\mu \nu } dx^\mu   \wedge dx ^\nu
+\V_{\mu }^1dx^\m    +  \V_{\mu }^{-1}dx^\m     +\M^{2} + \M ^{-2}  +\N  ^{
} .\cr }  }
It indicates  that an ``on-shell''      two-form in six dimensions, with kinetic
energy $|\pa B|^2$,   counts for $C^2_4=6 $ degrees of freedom.
($6=15-2\cdot 6+3\cdot 1$.) Once the  fields \field\ are introduced, one
combines the BRST operator  $Q$ of  section 2.3 with the    BRST operator
for the ordinary gauge symmetries  of two-forms. The latter involves the fields
in \field. This gives a   combined  nilpotent topological BRST operator
$\lies $,
defined as:
\eqn\syms{\eqalign{& \lies B_{\mu \nu} = \Psi_{\mu\nu} +
\pa_{[\m}\V_{\n]}
\quad
\lies \V_\m=-\Phi_\m+\pa_\m \M
\cr
& \lies \Psi_{\mu\nu} =\pa_{[\m}\Phi_{\n]}
\quad\quad\quad\quad
\lies \Phi_{\mu  } = \pa_{\mu } \Phi
\cr
&  \lies \Phi=0
\quad\quad\quad\quad\quad\quad \quad\lies  \M= \Phi.
\cr
}  }
The possibility of having a geometrical interpretation for the two-form and
its ghosts, briefly mentionned in the introduction, relies on the following
unifying equation:
\eqn\aaa {\eqalign{ (d+\lies)(B_{\mu \nu } dx^\mu   \wedge dx ^\nu
+\V_{\mu }^1dx^\m         +\M^{2} )=G_3
+\Psi_{\m\n}^1 dx^\mu   \wedge dx ^\nu
+\Phi_{\m }^2 dx^\mu +\Phi ^3,}}
\eqn\aaaa{ \eqalign{
 (d+\lies)( G_3
+\Psi_{\m\n}^1 dx^\mu   \wedge dx ^\nu
+\Phi_{\m }^2 dx^\mu   +\Phi ^3)=0
.  }  }

For the antighosts, one has:
\eqn\symsbar{\eqalign{
& \lies  \chi^{-1}_{\mu \nu\r} = \H_{\mu\nu\r} \quad\quad\quad \lies
\H_{\mu\nu\r} =0
\quad\quad\quad\lies  \Phi_{\mu  }^{-2} = \eta^{-1}_\m  \quad\quad\quad
\quad \lies \eta^{-1}_\m =0
 \cr
& \lies  \Phi^{-3}  = \eta^{-2} \quad \quad\quad\quad\lies  \eta^{-2} =0
 \quad\quad\quad\lies  \X^{ 1}  = \eta^{2}
\quad\quad \quad\quad\quad\lies  \eta^{2} =0
 \cr
}  }
\eqn\symsss{\eqalign{   \lies   \V_\m^{-1}=b_\m  ^{-1}\quad\quad   \lies b_\m=0
\quad\quad  \lies   \M ^{-2} = \b^{-1}
\quad\quad \lies   \b^{-1}=0
\quad\quad \lies  \N  = \b^1
\quad\quad \lies \b^{1}=0 . \cr
}  }

The gauge-fixing of the ordinary gauge invariance of the two-form gauge
field $B_{\m\n}$ and of its ghosts is   realized by the following action,
which must be added to the action \symQfcc:
\eqn\symsf{\eqalign{ I_{gf}=\int
\lies
 \big(  \V_\m ^{-1} \  (\pa_\n  B_{[\m\n]} +\pa_\m \N +\demi b_\m)
+  \M ^{-2}\pa_\m \V_\m^1  -cc\big).
\cr}  }
After expansion and elimination  of fields with algebraic equations of
motion, one gets:
\eqn\symsff{\eqalign{ I_{gf}\sim\int   &
|\pa_\n B_{[\m\n]}|^2 +|\pa_\m \N ^0|^2 +  \M ^{-2}\pa_\m\pa^\m \M^{ 2}
\cr &
- \V^{-1}_\m\pa_{[\m}\pa_{\n]} V^{1}_\n
+  \beta ^{-1}\pa_\m \V^{1}_\m +\b ^{ 1}\pa_\m   \V^{-1}_\m\ -
\V^{-1}_\m\pa_\n\Psi^{ 1}_{[\m\n]}\ ) -cc.
\cr}  }

By construction, the complete and fully gauge-fixed action
$ I_{ top}+I_{gf}$    is   invariant under the action of $\lies$.
 The last term in  \symsff\ can be absorbed into a redefinition
$\eta^{-1}_\m \to \eta^{-1}_\m +  \V_\m^{-1}$ in \symQfcc, and then
$\b^1\to\b^1+\X^1$.

 \newsec{Hamiltonian presentation}

To establish a link with the earliest introduction of the Yang--Mills TQFT
by Witten
\wittentopo, we observe that part of the Hamiltonian can be written as
$H=\demi \big\{Q,\bar Q \big\}
$,
with
\eqn\ham{\eqalign{ &
Q=\exp -i V  \big (
\int d^5x  \Psi_{ij}^{ 1}(x)  {{\delta}\over {\delta  B_{ij}(x)}} -cc \big )
\exp iV  -cc
\cr
}  }
\eqn\ham{\eqalign{ &
\bar Q=\exp  i V \big (
\int d^5x   \Psi_{ij}^{- 1}(x)  {{\delta}\over {\delta  B_{ij}(x)}}-cc \big )
\exp -iV
\cr
}  }
\eqn\ham{\eqalign{ &
  V=\int d^5x  \e ^{ijklm} B_{ij}(x) \pa_k {^cB}_{lm}(x).
   \cr
}  }
Here, all indices $i,j$ are five-dimensional. The field   $
\Psi_{ij}^{-1}(x)$ has 10 components
and is replaced by the
self-dual three-form $   \chi_{\m\n\r}^{-1}(x)$ in six dimensions. Of course,
this suggests that there is an interesting five-dimensional action, $\int
{^cB_2}\wedge
dB_2$, which should be independently quantized, perhaps by generalizing the
method given in \nekrasov\ for higher-dimensional Yang--Mills Chern-Simons
theories \rabin. Notice that there is the  possibility of adding    to
${^cB_2}\wedge
dB_2$ a   Yang--Mills- and/or gravitational  Chern--Simons- term,  since
$B_2$ can be defined as a   gauge
field two-form, as noted in the introduction  \simons.

\newsec{Interacting  cases. }
\subsec{Interactions with a Yang--Mills theory through a Chern-Simons coupling.}

   Coupling with    a Yang--Mills theory    can be introduced  by adding
by adding a Chern--Simons term  to the Abelian curvature of
the two-form, and  a Yang--Mills term $\int _6 {\Tr} |F_{\m\n}|^2$ to the
action of the two-forms that we will shortly determine. One defines
\eqn\symcs{\eqalign{
G_3=dB_2 +\lambda {\Tr} (A\wedge d A+{2\over 3}A\wedge A \wedge A),\cr
}  }
where $\lambda$ is a constant, possibly   adjusted to provide    theories with
anomaly-compensating mechanisms.

One modifies $Q$ in \syms\ as follows (\symsbar\ and \symsss\ remain the same)
\eqn\symsym{\eqalign{ & \lies B_{\mu \nu} = \Psi_{\mu\nu}^{1} +
\pa_{[\m}\V^{1}_{\n]}
+\lambda {\Tr} (c \pa_{[\m} A_{\n]})
 \quad\quad \lies \Psi^{1}_{\mu\nu} =\pa_{[\m}\Phi^{2}_{\n]}
 \quad\quad
\lies \Phi^{2}_{\mu  } = \pa_{\mu } \Phi^{3}
 \quad\quad  \lies \Phi^{3}=0
\cr &
\lies \V^{1}_\m=-\Phi^{2}_\m+\pa_\m \M^{2}
-\lambda {\Tr}( cc A_\m)\quad\quad \quad\lies  \M^{2}= \Phi^{3}
-{\lambda \over 3}{\Tr} (ccc).
\cr
&\lies  A_\mu=\pa_\mu c+[A_\m,c]
\quad\quad\quad\quad\quad\quad\quad\quad\quad  \lies c=-cc.
\cr}  }
One still has $\lies ^2=0$. One must complete the BRST  equations by $\lies
\bar c =b,  \lies  b=0$ for the sake of gauge-fixing the  ordinary
Yang--Mills symmetry.
The action   is  as in  \lsymQf, except that
in the gauge function \self,  $F_3=dB_2$ is   replaced
by $G_3 $.
(Notice that $\lies G_3=d \Psi $.)  Generalizing what we did  in section 2,
we get  the following $Q$-invariant action

  \eqn\symQfchs{\eqalign{ I_{\top} \sim \int   & {\Tr} |F_{\m\n}|^2+
{1\over \a}|\pa_{[\mu }B_{\nu\rho]} +\lambda   {\Tr} ( A{[_\m}\pa_\n
A_{\r]}+ {2\over 3} A{[_\m } A_\n A_{\r]})|^2\cr
&
-\chi^{\m\n\r}
\pa_{[\mu} \Psi^1_{\nu\rho]}
+\eta_\m ^{-1}\pa_\n\Psi_{[\m\n]}^{1}
-\X^1\pa_\m\eta_\m^{-1}\cr
&
+  \Phi^{-2}_\m \pa_{ \n}\pa_{[\m }\Phi^{ 2}_{\n]}
+\pa_\n   \Phi_\n^{-2}\pa_\m \Phi^{ 2}_\m
-  \Phi^{-3}\pa_\m\pa^\m \Phi^{ 3} -cc.
\cr
} }
This action     generalizes \symQfcc\ and contains   interactions between
$A_\m$,   and $B_{\m\n}$.
 As for the link to Poincar\'e supersymmetry, which holds true in the free
case $\lambda =0$, we do not know whether it is
compatible with  the interacting terms,   for the transformation laws and
for    the action.

\subsec{ Coupling to a Yang--Mills TQFT.}

In six dimensions, there is a Higgs Yang--Mills  TQFT   \bks. Rougly
speaking, this
theory is a dimensional reduction of the eight-dimensional pure Yang--Mills
TQFT  with  octonionic    self-duality equations as gauge functions, for
which  the topological BRST symmetry
 is also related to
Poincar\'e supersymmetry. Its action can be
substituted to
${\Tr} |F_{\m\n}|^2$ in \symQfchs. One must    modify \symsym\   to take into
account the transformation law of the Yang--Mills field as given in \bks.
This  gives new ghost interactions which modifies   \symQfchs.

 \newsec{Higher Dimensions than six.}

Here we consider the TQFT for a self-dual three-form in eight dimensions.
Formula \fieldtopr\ becomes:
 \eqn\fieldtopre{\eqalign{& F_{\mu \nu\rho\sigma} dx^\mu   \wedge dx ^\nu
\wedge dx ^ \rho \wedge dx ^ \sigma+\Psi_{\mu\nu\r}^1dx^\m
 \wedge dx ^\n \wedge dx ^\r +\chi^{-1}_{\mu \nu\rho\sigma}
dx^\mu   \wedge dx ^\nu   \wedge dx ^ \rho dx ^ \sigma
\cr &
+\Phi_{\m\n}^{2}dx^\m\wedge dx^\n+ \Phi_{\m\n}^{-2}dx^\m \wedge dx^\n
+\Phi_\m^{3}dx^\m+ \Phi_\m^{1}dx^\m
+ \Phi_\m^{-3}dx^\m
 +\Phi ^{4} +\Phi ^{2}  + \Phi
^{-2}+  \Phi ^{-4}.
\cr}  }
  $\chi^{-1}_{\mu \nu\rho\sigma}
 $ is a self-dual four-form antighost adapted for the eight-dimensional
self-duality condition of the curvature of $B_3$. Notice that there is no
need to double the degrees of freedom as in six dimensions, since a single
three-form determines the invariant $\int_8 dB_3\wedge dB_3$. By
generalizing what we
did in six dimensions, we obtain   fermionic Lagrange multipliers fields
that  are a two-form
$\eta_{\m\n}^{-1}$ and  three zero-forms $\eta ^{\pm 3}$ and  $\eta ^{3}$.
Eventually,      the TQFT multiplet \mul\ becomes
\eqn\mule {\eqalign{(B_{\m\n\r}, \Phi_{\m \n  }^{\pm 2},
\Phi ^{\pm 4}, \Phi  ^{\pm 2},
 \Psi_{\m \n\r }^{1}, \chi_{\mu
\nu\r\sigma}^{-1},
\eta_{\mu \n }^{-1} ,\Phi_\m^{\pm 3},    \Phi_\m^{1},
\eta^{\pm 3}, \eta^{1})
\cr }  }
The number of degrees of freedom of the TQFT, in addition to that of the 3-form
$B_{\m\n\r}$, is $2\cdot 28+4=60$ for the bosons and
$56+35+28+3\cdot 8+3=146$ for the  fermions.
This set of fields can be twisted into the D=8 supermultiplet  of
a three-form made of  two two-forms, two one-forms, two zero-forms, and one
gravitino
and three spinors
\eqn\suse {\eqalign{(B_{\m\n\r}, 2A_{\m \n  },
2A_{\m    }, 2\Phi, 1\Psi_\m^\a, 3\kappa^\a)
\cr }  }
This $N=1$ multiplet has on-shell states that are given by
tensor product of   the on-shell states of the $N=1$
Yang--Mills multiplet   with the $4+\bar 4$ of the
little group $SO(6)$.
Let us indeed check how the counting of degrees of freedom of the
supersymmetric multiplet \suse\ matches that of the
TQFT in \mule. When counting the number of degrees of freedom
we must as usual subtract the ones which are pure gauges.
  The number of
boson  degrees of freedom  is thus   $7\cdot6=42$ for both two-forms,
$7\cdot 2=14$ for the two one-forms    and $2$ for   both zero-forms, which
makes  $58$ boson  degrees of freedom to be added to those of   the
three-form. For the  gravitino and   three spinors, the spinor  index  $\a$
runs over 16 values. Thus we have $  6\cdot 16=96$ degrees of
freedom for the gravitino and $3\cdot 16=48$ ones for the three spinors,
that is
$144=96+48$ fermionic  degrees of freedom.
If, as in the six-dimensional case, we eliminate a pair of bosons against a
pair of fermions    in  \mule, for instance $\Phi^{\pm 4}$ against
$\eta^{\pm 3}$,
we get an exact
matching between the fields in the TQFT system \mule\ and the field
of the supersymmetric system \suse, with $58 $ bosons and $144$ fermions.
This  indicates the possibility of a twist
between the two  systems. The striking difference between the situations in six
and eight
dimensions is  the apparition of the gravitino in eight dimensions, which
gives a supersymmetric theory depending on a spin-$3/2$ field with no
gravity coupling. Again, interactions with Yang--Mills field can be
introduced by adding  Chern classes    in the self-duality equations, while
the Yang--Mills dynamics can be defined from  an eight-dimensional TQFT as
in \bks.

Preliminary computations  also suggest that the TQFT system in ten
dimensions for a
self-dual  four-form might be twistable into the supersymmetric multiplet
of type
IIB supergravity.

\newsec{Conclusion}

We have shown that the notion  of a
twist between TQFTs and supersymmetric theories can be generalized
from   that known to exist in two and four dimensions
provided one  eliminates  topological pairs. In particular, we have
considered a topological theory of  two-forms  in six
dimensions and established the  relation    of its
BRST symmetry to  the twisted transformations of the
tensor supermultiplet of $(2,0)$
  Poincar\'e  supersymmetry.
Although there are a number of features of the twisting procedure
that remain mysterious, it is apparent that
 there is a general relationship between
supersymmetry transformations and the BRST transformations
associated with topological theories. Central to this relationship
is the fact that  TQFTs
 involve   self-duality conditions  as their gauge-fixing conditions
and that self-duality conditions arise naturally in supersymmetric
theories as a consequence of the
 intimate relationship between
self-duality, supersymmetry  and spinor chirality. It is interesting
to note that the six and four dimensional supersymmetric theories
both contain spinor fields which satisfy a symplectic Majorana
condition.
\par
We have also presented interactive
theories   that could be interesting to explore. In eight dimensions,
there are indications that these features   repeat themselves for the
theory of three-forms, leading, after a twist, to a theory with a
gravitino, but no graviton.
\par
 An intriguing  open
question is to examine if another  expression of the   six-dimensional-TQFT
can be untwisted in the action recently proposed in \tonin\ for
self-dual two-forms.

\vskip 0.5cm

\noindent{\bf Acknowledgements}

We are grateful to E. Cremmer for most useful and  interesting discussions
  related to this work.

 \listrefs
\bye